\documentclass{PoS}
\usepackage{axodraw,epsfig,amsmath}
\newcommand{\bea}{\begin{eqnarray}}
\newcommand{\eea}{\end{eqnarray}}
\newcommand{\beq}{\begin{equation}}
\newcommand{\beqns}{\begin{eqnarray*}}
\newcommand{\eeqns}{\end{eqnarray*}}
\newcommand{\eeq}{\end{equation}}

\newcommand{\pdir}{p\kern -5.2pt\raise 0.2ex\hbox {/}}
\newcommand{\vdir}{v\kern -5.75pt\raise 0.15ex\hbox {/}}
\newcommand{\kdir}{k\kern -5.75pt\raise 0.15ex\hbox {/}}
\newcommand{\epsdir}{\epsilon\kern -5.0pt\raise 0.15ex\hbox {/}}
\newcommand{\bvdir}{\bar{v}\kern -5.75pt\raise 0.15ex\hbox {/}}
\newcommand{\Ddir}{D\kern -7.75pt\raise 0.20ex\hbox {/}}
\newcommand{\Adir}{A\kern -7.75pt\raise 0.20ex\hbox {/}}
\newcommand{\ldir}{l\kern -5.0pt\raise 0.2ex\hbox{/}}
\newcommand{\varepsdir}{\varepsilon\kern -5.5pt\raise 0.15ex\hbox{/}}

\newcommand{\msb}{\overline{\rm{MS}}}

\newcommand{\lgl}{\langle}
\newcommand{\rgl}{\rangle}

\def\elematrice#1#2#3{\lgl#1|#2|#3\rgl}

\def\Journal#1#2#3#4{{#1} {\bf #2}, #3 (#4)}

\def\NPB{{\em Nucl. Phys.} B}
\def\PLB{{\em Phys. Lett.}  B}
\def\PRL{\em Phys. Rev. Lett.}

\def\PRD{{\em Phys. Rev.} D}

\title{Twisted mass QCD in the charm sector}
\ShortTitle{Twisted mass QCD in the charm sector}
\author{\speaker{Beno\^\i t Blossier}\\
DESY, Platanenallee 6, 15738 Zeuthen, Germany\\
E-mail: \email{Benoit.Blossier@desy.de}}
\author{Gregorio Herdoiza\\
INFN -- Sezione di Roma Tor Vergata, Via della Ricerca Scientifica 1, 
I-00133 Roma, Italy}
\author{Silvano Simula\\
INFN -- Sezione di Roma Tre, Via della Vasca Navale 84, I-00146 Roma, Italy}

\author{for the European Twisted Mass Collaboration (ETMC)}

\abstract{We present preliminary results for the charm quark mass $m_c$ 
and the $D$ and $D_s$ mesons decay constants $f_D$ and 
$f_{D_s}$ from a lattice QCD calculation with ${\rm N_f}$ = 2
dynamical fermions. We use the twisted mass fermionic action defined at maximal twist so that 
physical quantities are automatically ${\cal O}(a)$ improved. Two lattice spacings are considered. The 
charm quark mass has been
renormalised in the RI-MOM scheme. After a matching to the 
$\msb$ scheme, we obtain from the simulation
at a fine lattice ($a \sim 0.09$ fm) 
$m_c^{\msb}(m_c) = 1.481 \pm 0.022 \pm 0.092$ GeV, 
$f_D = 205 \pm 13 \pm 17$ MeV, 
$f_{D_s} = 271 \pm 6 \pm 6$ MeV 
and from the simulation
at the finer lattice ($a \sim 0.07$ fm) 
$m_c^{\msb}(m_c) = 1.474 \pm 0.041 \pm 0.132$ GeV,  
$f_D = 230 \pm 31 \pm 8$ MeV and 
$f_{D_s} = 264 \pm 5 \pm 8$ MeV. 
We chose three renormalisation conditions to determine $m_c$: the spread
between the final results contributes to the systematic error. At both lattice spacings, 
particularly at the finer one, the error on 
$m_c$ is dominated by present uncertainty on the renormalisation constant $Z_P$, 
which should be reduced
before 
performing a reliable continuum limit. 
}
\FullConference{XXV International Symposium on Lattice Field Theory\\
                            
                            30 July - 4 August 2007\\ 
                            
                            University of Regensburg, Germany}

\begin{document}
\section{Introduction}

The physics of charm bound states regained recently the attention of particle physicists with the discovery 
of the new resonances $X(3872)$, $X(3943)$, $Y(3940)$, $Y(4260)$ 
and $Z(3930)$ \cite{Xstatecharmonium} and of a very narrow scalar state $D_s(2317)$ \cite{Dsnarrow}, 
whose composition is
still an open question. The experimental evidence for oscillations in the $D^0 - \overline{D^0}$ system 
\cite{DDbarmix} might be the first signal for physics beyond the Standard Model in the charm sector. 
Even if it is expected that the long-distance physics is a dominant effect in that
process, the $\Delta C = 2$ contribution to $x_D=\Delta M_D/\Gamma_D$ might be not
negligible. It is given by a 
box diagram, 
as in the $B-\overline{B}$ system, and it is proportional to $f^2_D$.
In the CKM matrix, $V_{cs}$ is one of the elements having the largest uncertainty when one does not impose the 
$3 \times 3$ unitarity:  $\frac{\Delta |V_{cs}|}{|V_{cs}|}=9.82$\% \cite{PDG}: 
most of it comes from the theory. An appropriate way to 
extract $V_{cs}$ is to measure the leptonic decay width $D_s \to l \bar{\nu}$, 
which however requires the estimation of the decay constant $f_{D_s}$. Eventually it is well established that 
$V_{cb}$
can be constrained by analysing the inclusive semileptonic decay $\overline{B} \to X_c l \bar{\nu}$. An OPE
is used in the Heavy Quark Expansion (HQE) and the total branching ratio reads \cite{BRSLmc}
${\rm BR}_{\rm SL}(\overline{B}) = {\rm BR}^{(0)}_{\rm SL}\, [1+f(r)],\, 
r=\left(\frac{m_c}{m_b}\right)^2,
$
where ${\rm BR}^{(0)}_{\rm SL}$ is the dominant term in the $m_c/m_b$ expansion.\\
From those phenomenological aspects we conclude that it is important to determine as precisely as 
possible the decay constants $f_D$, $f_{D_s}$ and the charm quark mass $m_c$. 
That quark is of course too heavy 
to make predictions by using Chiral Perturbation Theory ($\chi$PT) and too 
light to compute the 
amplitudes by using only the HQE: corrections of 
${\cal O}(\Lambda_{QCD}/m_c)^n$ and ${\cal O}(m_c/m_b)^n$ 
might be larger than the precision of few \% that we want to reach.\\ 
Lattice QCD is a good tool to study the charm sector. However
the continuum limit can be difficult to reach because of large cut off 
effects (typically, $0.2 < am^0_c < 0.4$). 
It is therefore crucial to improve the action and the currents regularised on the lattice. 
Several theories
proposed in the literature so far have the
common property that they require the tuning of a certain number of 
parameters to achieve improvement, by applying
the Symanzik's program \cite{Symanzik}. On the other side it has been shown \cite{tmQCDsymmetries} 
that Twisted mass QCD (TmQCD) 
\cite{AokitmQCD,FGSW} requires the tuning of a single parameter (the untwisted mass $m_0$),
so that hadronic quantities like the pseudoscalar meson masses and decay constants are 
automatically ${\cal O}(a)$ improved at maximal twist. Other nice properties of such an action 
are that the physical quark mass  is
related to the twisted mass parameter of the action, its renormalisation is only
multiplicative and the pseudoscalar decay constant does not require the introduction of 
any
renormalisation constant \cite{FGSW}: for two quark flavors 1 and 2 (for example a light
flavor $\ell$ and a heavy flavor $h$) it is simply given by
\beq\label{fps}
f_{PS}(\mu_1, \mu_2) = \frac{\mu_1+\mu_2}{m^2_{PS}(\mu_1, \mu_2)} 
|\elematrice{0}{P^c(0)}{P}|,\quad P^c=\bar{\psi}_1(r) \gamma^5 \psi_2(-r),
\eeq
where $r$ is the Wilson parameter 
and we define the composite operator $P^c$ in the physical basis.

We present a preliminary lattice QCD determination of the charm quark mass $m_c$ and the 
decay
constants $f_D$ and $f_{D_s}$.
We have performed full dynamical simulations for 
${\rm N_f}$ = 2 light degenerate sea quarks; the strange and the charm quarks
have been added in the valence sector.
\noindent The calculation is based on the analysis of the gauge configurations 
ensembles $B_1$ -- $B_5$ and $C_1$ -- $C_4$ \cite{ETMC} 
(240 and 130 configurations of $B_1$ -- $B_5$ and $C_1$ -- $C_4$, respectively, have been
analysed) 
which have been generated with the TlSym gauge action at $\beta=3.9$ ($a=0.0855(5)(31)$~fm) 
and $\beta=4.05$ 
($a = 0.0667(5)(24)$~fm) respectively and the twisted mass fermionic action defined at maximal twist. 
The light quark masses are in the range [$m_s/6$, $2m_s/3$], to perform the chiral extrapolation, and we used 
masses around the strange mass and the charm mass to do the appropriate interpolations.

\noindent At each sea quark mass we have computed the two-point correlation functions of 
pseudoscalar mesons. Each measurement has been separated by 20 HMC trajectories, which is enough to avoid 
autocorrelation time effects. The statistical accuracy 
has been improved by using all to all stochastic propagators. Statistical errors on the meson masses and decay
constants are evaluated at a given sea quark mass by using a jacknife procedure with 10 
measurements discarded in each bin.  
The error obtained after a combination of data coming from simulations with different sea 
quark masses (i.e. statistically independent) is computed using a bootstrap method. 
\section{Charm quark mass}
To estimate the charm quark mass we use the following strategy: we compute the 
pseudoscalar meson mass 
$m_{PS}(\mu_{\rm sea}, \mu_{\ell}, \mu_h)$ (where $\ell$ and $h$ are valence light and heavy quarks
respectively) 
at the points $\mu_{\ell} = \mu_{\rm sea}$, which allows us to extrapolate down to the 
physical light quark mass $\mu_{ud}\equiv \frac{\mu_u + \mu_d}{2}$,
previously determined in \cite{ETMCletter}. Once the dependence on the light quark 
mass 
has been taken into account, one studies the dependence of $m_{PS}$ on the heavy mass 
$\mu_h$. 
The bare charm 
quark mass $\mu_c$ is determined by using the following condition:
$m_{PS}(\mu_{ud},\mu_c)=m_D$. 

\begin{figure}
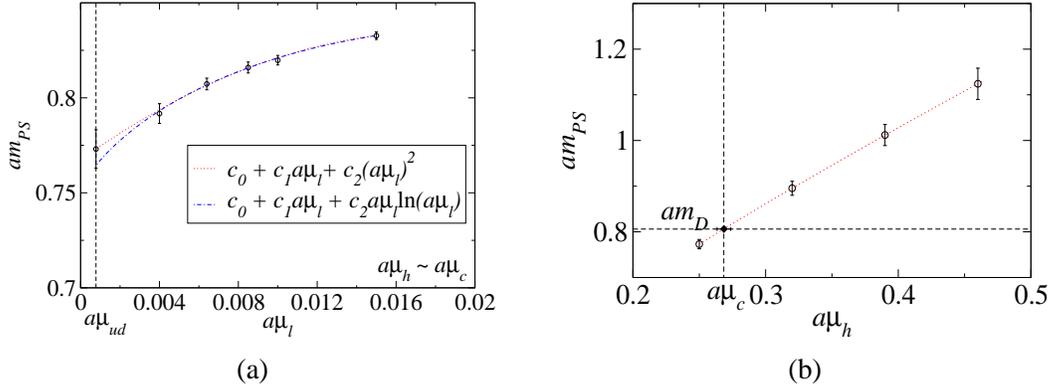

\begin{center}
\begin{tabular}{ccc}
\includegraphics*[width=6.5cm, height=4.5cm]{plotmPSextrsea.eps}
&&
\includegraphics*[width=6.5cm, height=4.5cm]{mpsheavynote.eps}\\
(a)&&(b)\\
\end{tabular}
\end{center}
\vspace{-0.5cm}
\caption{\label{fig1}
(a) Chiral extrapolation of the pseudoscalar heavy-light meson masses ($\beta=3.9$).
(b) Pseudoscalar heavy-light meson mass in function of the heavy quark mass ($\beta=3.9$).}
\end{figure}

\begin{figure}[t]
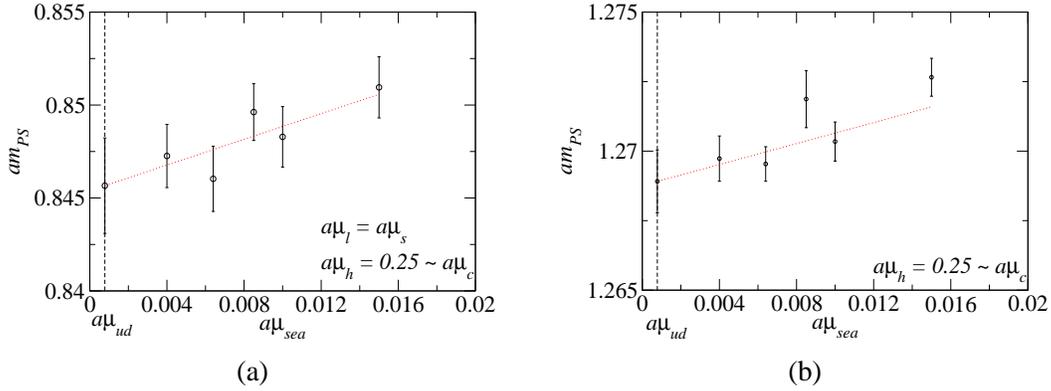

\begin{center}
\begin{tabular}{ccc}
\includegraphics*[width=6.5cm, height=4.5cm]{mpstrangeextrsea.eps}
&&
\includegraphics*[width=6.5cm, height=4.5cm]{metaextrsea.eps}\\
(a)&&(b)\\
\end{tabular}
\end{center}
\vspace{-0.5cm}
\caption{\label{figdsetac}
(a) Chiral extrapolation of $m_{PS}(\mu_{\rm sea}, \mu_{\ell}=m_s, \mu_h\sim m_c)$ ($\beta=3.9$).
(b) Chiral extrapolation of $m_{PS}(\mu_{\rm sea}, \mu_h, \mu_h)$ ($\beta=3.9$).}
\end{figure}

\noindent We have performed a quadratic extrapolation of $m_{PS}(\mu_{\ell}=\mu_{\rm sea},\mu_h)$ in $\mu_{\ell}$ 
down to 
$\mu_{\ell}=\mu_{ud}$. This is illustrated in Figure \ref{fig1} (left). 
We have also introduced a logarithmic dependence on $\mu_{\ell}$: 
$m_{PS}(\mu_{\ell},\mu_h) = c_0(\mu_h) + a\mu_{\ell} [c_1(\mu_h) + c_2 (\mu_h) 
\ln(a \mu_{\ell})]\,$.
As a third possibility we have done a simple linear extrapolation. The spread between these different fits is
included in the systematic error at the end of the computation.
\noindent We then performed a fit of $m_{PS}$ as a function of $\mu_h$: we used either a quadratic 
polynomial in
$\mu_h$, or a quadratic polynomial in $\frac{1}{\mu_h}$, or also, as a
third ansatz, $m_{PS}= d_0 + d_1 \mu_h + \frac{d_2}{\mu_h}$. We show in 
Figure \ref{fig1} (right) the
quality of the latter fit function, that appears to be the most appropriate to 
describe the data. Once the
bare charm quark mass $\mu_c$ is extracted, we renormalise 
it in the RI-MOM scheme: 
$m^{\rm RI-MOM}_c = Z^{\rm RI-MOM}_\mu \mu_c$ where $Z^{\rm RI-MOM}_\mu =1/Z^{\rm RI-MOM}_P$ in
TmQCD \cite{tmQCDsymmetries}. Finally we perform a matching onto the 
$\msb$ scheme.
We indicate in Table \ref{tab2} the value of $m_c^{\msb}(m_c)$ for the two lattice
spacings,
knowing that $Z^{\rm RI-MOM}_P(\beta=3.9,1/a) = 0.39(1)(2)$ and the \emph{preliminary} value of
$Z^{\rm RI-MOM}_P(\beta=4.05,1/a) = 0.40(1)(4)$\footnote{At this $\beta$ a chiral extrapolation has been
performed in the valence sector, at $a\mu_{\rm sea} = 0.003$, but not yet in the sea sector. 
However it was found at $\beta=3.9$ that $Z_P$ depends only weakly on the sea quark mass. 
Thus as a first
step of the analysis we will include the sea effects in the systematic
uncertainty. Moreover, an alternative estimate of $Z_P$ at $\beta=4.05$
may come from scaling as described in \cite{Dimopscaling}, which brings
our final estimate of the systematic error to 0.04. } \cite{ZDimopoulos}. The first 
error on $m_c$ is statistical, the second is the systematic error coming from $Z_P$, the third comes 
from the uncertainty on $a$ and the last one is the systematic error from the chiral extrapolation. In that table
we have also collected the result of two other determinations of $m_c$, by using the following renormalisation 
conditions:\\
1) $m_{PS}(\mu_{\rm sea} =\mu_{ud}, \mu_{\ell}=\mu_s,\mu_c)=m_{D_s}$ and 2) 
$m_{PS}(\mu_{\rm sea} =\mu_{ud},\mu_c,\mu_c)=m_{\eta_c}$.\\
$\mu_s$ is the bare strange quark mass which has been determined in \cite{msTarantino}. 
The heavy-heavy pseudoscalar meson correlator has been computed by using the interpolating field 
$\bar{\psi}_h(r)\gamma^5 \psi_h(r)$.\\ 
The dependence of those two observables on the sea quark mass is very weak, as shown in Figure 
\ref{figdsetac}. We used a linear fit in $\mu_{\ell}$ and $\mu_{\rm sea}$ to interpolate to $\mu_s$ 
and to extrapolate down to $\mu_{ud}$ respectively:
\vspace{-0.6cm}
\bea\nonumber
m_{PS}(\mu_{\rm sea},\mu_{\ell},\mu_h)& =& p_0(\mu_h) + 
a \mu_{\ell} p_1(\mu_h) + 
a\mu_{\rm sea} [p_2(\mu_h) + a \mu_{\ell} p_3(\mu_h)]\,,\\
\nonumber
m_{PS}(\mu_{\rm sea}, \mu_h, \mu_h)& = &q_0(\mu_h) + a \mu_{\rm sea} 
q_1(\mu_h).
\eea
\vspace{-1.0cm}

\noindent The uncertainty coming from the chiral extrapolation is quite reduced compared to 
$m_{PS}(\mu_{\rm sea}=\mu_{\ell},\mu_h)$. However we introduce a small 
uncertainty from $\mu_s$ on the first observable (third error on
$m_c^{\msb}(m_c,m_{D_s})$) 
and from the disconnected diagram which contributes to the second but that we did not compute. The last error
on $m_c^{\msb}(m_c,m_{D_s})$ and $m_c^{\msb}(m_c,m_{\eta_c})$ comes from the uncertainty on $a$.\\ 
We have not performed a continuum limit extrapolation yet because only 2 lattice spacings have been 
considered so far and the uncertainty on $Z_P$ is still rather large, especially at
$\beta=4.05$.
Concerning cut off effects, it is remarkable that they appear to be rather weak on the 
unrenormalised charm mass at 
$\beta=4.05$: 
indeed, the value of $\mu_c$
extracted from the 3 observables are much closer at this $\beta$ than at $\beta=3.9$.

\begin{table}[b]
\begin{center}
\begin{tabular}{|c|c|c|c|c|c|}
\hline
$\beta$&$m^{\msb}_c(m_c,m_D)$&$m^{\msb}_c(m_c,m_{D_s})$&$m^{\msb}_c(m_c,m_{\eta_c})$\\
\hline
3.9&1.481(22)(63)(8)(27) GeV&1.450(12)(61)(10)(15) GeV&1.420(5)(60)(6) GeV\\
4.05&1.474(41)(129)(15)(5) GeV&1.498(6)(130)(12)(18) GeV&1.479(2)(129)(8) GeV\\
\hline
\end{tabular}
\end{center}
\caption{\label{tab2} Charm quark mass fixed by using different observables.}
\end{table}

\noindent To conclude this section we note that our
values of $m_c$ are large with respect to most of the recent lattice estimations 
\cite{latticemc}: 
however the currently large uncertainty on $Z_P$ at $\beta=4.05$ implies that 
\emph{any conclusion about the continuum limit result would be untimely}. 
\section{Heavy-light meson decay constants}
\begin{figure}[t]
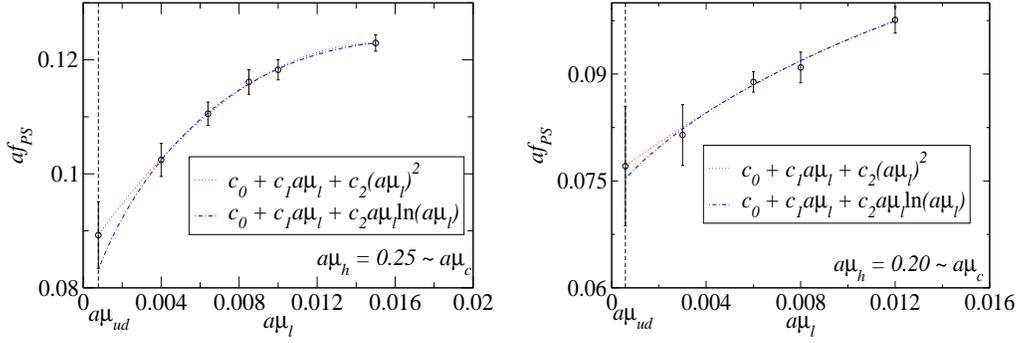

\begin{center}
\begin{tabular}{cc}
\includegraphics*[width=6.5cm, height=4.5cm]{fpextrsea.eps}
&
\includegraphics*[width=6.5cm, height=4.5cm]{fdseaextrb405.eps}
\end{tabular}
\end{center}
\vspace{-0.5cm}
\caption{\label{figfdchiral} Chiral extrapolation of $f_{PS}(\mu_{\ell}=\mu_{\rm sea}, \mu_h\sim \mu_c)$ at $\beta=3.9$
(left) and $\beta=4.05$ (right).}
\end{figure}

To determine $f_D$ and $f_{D_s}$ we employ the same strategy as in the previous section, 
using eq.~(\ref{fps}). We show 
in Figure \ref{figfdchiral} 
the chiral extrapolation of $f_{PS}(\mu_{\rm sea}=\mu_{\ell}, \mu_h)$ at $\mu_h\sim \mu_c$ down to the 
physical light quark mass. We found 
that introducing a quadratic term in the extrapolation improves the fit, particularly for the 
coarse lattice. As before, we also introduced also a logarithmic dependence 
on $\mu_{\ell}$ in the fit. We include
the spread between the different chiral extrapolations in the systematic error.\\
\noindent At $\mu_{\ell} \sim \mu_s$, $f_{PS}(\mu_{\rm sea}, \mu_{\ell}, \mu_h)$ has 
a similar linear dependence on $\mu_{\ell}$ to the one of $m_{PS}(\mu_{\rm sea}, \mu_{\ell}, \mu_h)$. 
Moreover for both the dependence on $\mu_{\rm sea}$ is weak and very well described by a linear
fit as well. 
\begin{table}[b]
\begin{center}
\begin{tabular}{|c|c|c|}
\hline
$\beta$&3.9&4.05\\
\hline
$f_D$&205(13)(3)(17) MeV&230(31)(6)(6) MeV\\
$f_{D_s}$&271(6)(4)(5) MeV&264(5)(4)(7)\\
$\frac{f_{D_s}}{f_D}$&1.35(4)(1)(7)&1.13(28)(2)(2)\\
\hline
\end{tabular}
\end{center}
\caption{\label{tab3} Decay constants $f_D$ and $f_{D_s}$
and $f_{D_s}/f_D$ from our simulation}
\end{table}

\begin{figure}[t]
\begin{center}
\includegraphics*[width=6.5cm, height=4.5cm]{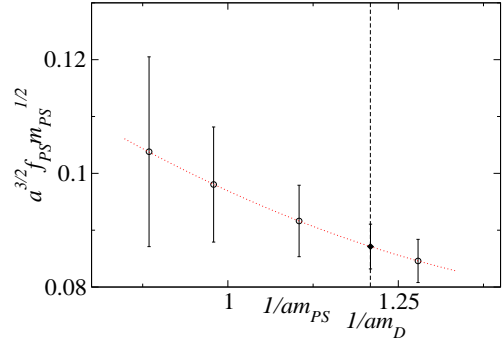}
\end{center}
\vspace{-0.8cm}
\caption{\label{fig4} Scaling law of $f_{PS} \sqrt{m_{PS}}(\mu_{ud},\mu_h)$ as a function of 
$1/m_{PS}(\mu_{ud},\mu_h)$ at $\beta=3.9$.}
\end{figure}

\noindent We performed a fit of $f_{PS}\sqrt{m_{PS}}$ with a quadratic polynomial in 
$\frac{1}{m_{PS}}$ (we can not isolate any logarithmic dependence on $\alpha_s(m_{PS})$ in our range
of heavy masses):
$f_{PS}\sqrt{m_{PS}}
=f_0 +\frac{f_1}{m_{PS}}+\frac{f_2}{m^2_{PS}}\, .
$
We show in Figure \ref{fig4} the quality of the fit for $f_{PS}\sqrt{m_{PS}} (\mu_{ud},\mu_h)$.

\begin{figure}
\begin{center}
\begin{tabular}{cc}
\begin{picture}(140,-20)(140,-20)
\LinAxis(0,-10)(120,-10)(4,4,-3.2,0,0.2)
\LinAxis(0,-200)(120,-200)(4,4,3.2,0,0.2)
\Line(0,-200)(0,-10)
\Line(120,-200)(120,-10)
\DashLine(0,-30)(120,-30){2}

\Text(0,-210)[]{\small{180}}
\Text(30,-210)[]{\small{200}}
\Text(60,-210)[]{\small{220}}
\Text(90,-210)[]{\small{240}}
\Text(120,-210)[]{\small{260}}
\Text(60,-230)[]{\small{$f_D$ (MeV)}}

\SetColor{Black}
\Line(36,-20)(92,-20)
\Line(36,-22)(36,-18)
\Line(92,-22)(92,-18)
\CCirc(64,-20){3}{Turquoise}{Turquoise}

\SetColor{Black}
\Line(4,-80)(68,-80)
\Line(4,-82)(4,-78)
\Line(68,-82)(68,-78)
\CCirc(36,-80){3}{Red}{Red}

\SetColor{Black}
\Line(32,-100)(96,-100)
\Line(32,-102)(32,-98)
\Line(96,-102)(96,-98)
\CCirc(64,-100){3}{Blue}{Blue}

\SetColor{Black}
\Line(30,-120)(85,-120)
\Line(30,-122)(30,-118)
\Line(85,-122)(85,-118)
\CCirc(50,-120){3}{Blue}{Blue}

\SetColor{Black}
\Line(16,-140)(88,-140)
\Line(16,-142)(16,-138)
\Line(88,-142)(88,-138)
\CCirc(52,-140){3}{Blue}{Blue}

\SetColor{Black}
\Line(4,-160)(56,-160)
\Line(4,-162)(4,-158)
\Line(56,-162)(56,-158)
\CCirc(30,-160){3}{Green}{Green}

\SetColor{Black}
\Line(40,-180)(46,-180)
\Line(40,-182)(40,-178)
\Line(46,-182)(46,-178)
\CCirc(43,-180){3}{Green}{Green}

\Text(140,-20)[l]{\small{CLEO-c, 2005}}
\SetColor{Red}
\Text(140,-40)[l]{\small{Alpha, ${\rm N_f}=0$, 2003}}
\Text(140,-60)[l]{\small{Rome 2, ${\rm N_f}=0$, 2003}}
\Text(140,-80)[l]{\small{QCDSF, ${\rm N_f}=0$, 2007}}
\SetColor{Blue}
\Text(140,-100)[l]{\small{CP-PACS, ${\rm N_f}=2$, 2000}}
\Text(140,-120)[l]{\small{MILC, ${\rm N_f}=2$, 2002}}
\Text(140,-140)[l]{\small{ETMC, ${\rm N_f}=2$, 2007}}
\SetColor{Green}
\Text(140,-160)[l]{\small{FNAL/MILC, ${\rm N_f}=2+1$, 2005}}
\Text(140,-180)[l]{\small{HPQCD/UKQCD, ${\rm N_f}=2+1$, 2007}}
\end{picture}
&
\begin{picture}(-10,-20)(-10,-20)
\LinAxis(0,-10)(120,-10)(4,4,-3.2,0,0.2)
\LinAxis(0,-200)(120,-200)(4,4,3.2,0,0.2)
\Line(0,-200)(0,-10)
\Line(120,-200)(120,-10)
\DashLine(0,-30)(120,-30){2}

\Text(0,-210)[]{\small{200}}
\Text(30,-210)[]{\small{225}}
\Text(60,-210)[]{\small{250}}
\Text(90,-210)[]{\small{275}}
\Text(120,-210)[]{\small{300}}
\Text(60,-230)[]{\small{$f_{D_s}$ (MeV)}}

\SetColor{Black}
\Line(73,-20)(103,-20)
\Line(73,-22)(73,-18)
\Line(103,-22)(103,-18)
\CCirc(88,-20){3}{Turquoise}{Turquoise}

\SetColor{Black}
\Line(51,-40)(71,-40)
\Line(51,-42)(51,-38)
\Line(71,-42)(71,-38)
\CCirc(61,-40){3}{Red}{Red}

\SetColor{Black}
\Line(40,-60)(56,-60)
\Line(40,-62)(40,-58)
\Line(56,-62)(56,-58)
\CCirc(48,-60){3}{Red}{Red}

\SetColor{Black}
\Line(12,-80)(40,-80)
\Line(12,-82)(12,-78)
\Line(40,-82)(40,-78)
\CCirc(26,-80){3}{Red}{Red}

\SetColor{Black}
\Line(58,-100)(105,-100)
\Line(58,-102)(58,-98)
\Line(105,-102)(105,-98)
\CCirc(80,-100){3}{Blue}{Blue}

\SetColor{Black}
\Line(19,-120)(85,-120)
\Line(19,-122)(19,-118)
\Line(85,-122)(85,-118)
\CCirc(51,-120){3}{Blue}{Blue}

\SetColor{Black}
\Line(88,-140)(68,-140)
\Line(88,-142)(88,-138)
\Line(68,-142)(68,-138)
\CCirc(78,-140){3}{Blue}{Blue}

\SetColor{Black}
\Line(34,-160)(78,-160)
\Line(34,-162)(34,-158)
\Line(78,-162)(78,-158)
\CCirc(59,-160){3}{Green}{Green}

\SetColor{Black}
\Line(47,-180)(55,-180)
\Line(47,-182)(47,-178)
\Line(55,-182)(55,-178)
\CCirc(51,-180){3}{Green}{Green}

\end{picture}\\
\end{tabular}
\end{center}
\vspace{7cm}
\caption{\label{figfdsummary} Recent lattice computations and experimental measurements of the 
decay constants $f_D$ (left panel) and 
$f_{D_s}$ (right panel). The ETMC values that we indicate are obtained by doing an average of the 
data at $\beta=3.9$ and $\beta=4.05$. 
}
\end{figure}
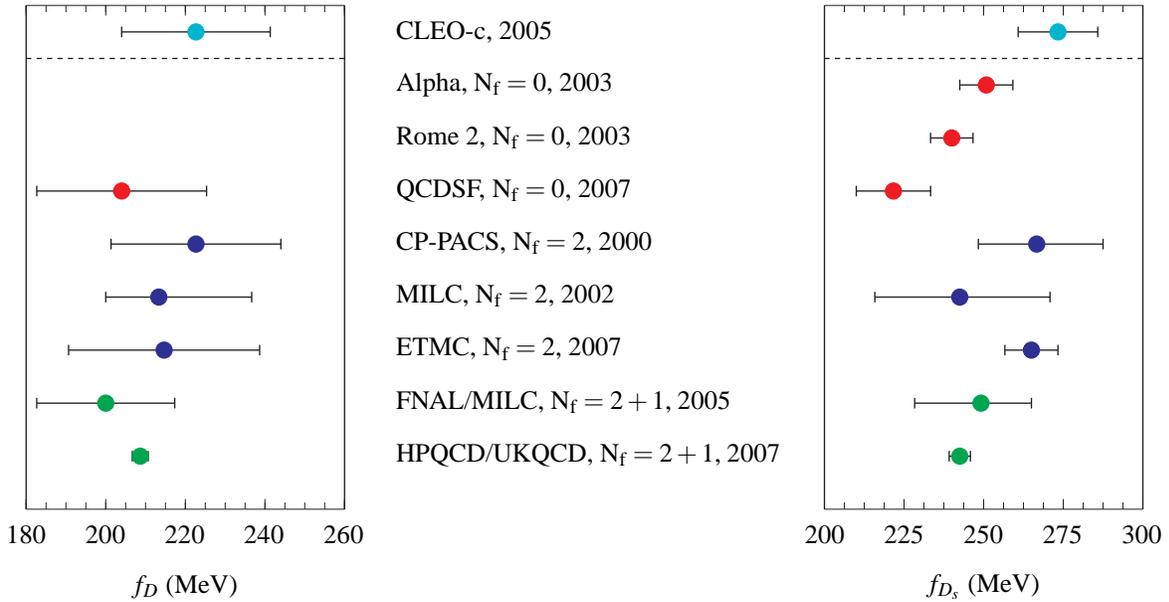

\noindent We give in Table \ref{tab3} our values of $f_D$, $f_{D_s}$ 
and $f_{D_s}/f_D$. The first error is a statistical error, the second error on $f_{D_s}$ 
and on $f_{D_s}/f_D$
comes from the uncertainty on
the bare strange quark mass, the second error on $f_D$ and the third error on $f_{D_s}$ 
come from the uncertainty on the lattice spacing. The last error on $f_D$ and 
$f_{D_s}/f_D$ comes from the spread 
between different
chiral fits.
We have 
collected the most recent lattice estimations of those quantities in Figure \ref{figfdsummary} 
\cite{latticefd}. 
On the experimental side CLEO-c measured $f_D=222.6\pm 16.7 ^{+2.8}_{-3.4}$ MeV \cite{CLEOfd}, 
$f_{D_s}=274\pm 13 \pm 7 $ MeV \cite{CLEOfds}, which is a 
combination of analysis of $D_s \to \mu$ and $D_s \to \tau$ leptonic decays.
Note that BABAR measured $f_{D_s}=283 \pm 17 \pm 7 \pm 14$ MeV \cite{BABARfds}.

\section{Summary}
We have presented preliminary results of a lattice computation of the charm quark mass $m_c$ and the $D$ and 
$D_s$ mesons decay constants $f_D$ and $f_{D_s}$ by using the ${\rm N_f}=2$ TmQCD action defined at maximal twist.
Encouraging results are found concerning cut off effects. Indeed, the bare charm quark mass extracted from 3
different observables looks consistent at the finer lattice. 
However, before performing the continuum limit
on $m_c$, $f_D$ and $f_{D_s}$, we still have to reduce as much as possible 
the uncertainty on the renormalisation 
constant $Z_P$ and to increase the statistics at $\beta=4.05$.
A more detailed study of the light quark dependence of the $D$ mesons masses and decay constants, 
based on heavy-light chiral perturbation theory, is still missing at this stage. Finally a better control on the continuum limit 
extrapolation will come from the on-going analysis of the data at a coarser lattice ($a \sim 0.1$ fm).


\begin{thebibliography}{}
\bibitem{Xstatecharmonium} 
S.~K.~Choi \emph{et al} (Belle Collaboration), 
\Journal{\PRL}{91}{262001}{2003}; S.~K.~Choi \emph{et al} (Belle Collaboration), 
\Journal{\PRL}{94}{182002}{2005}; B.~Aubert \emph{et al} (BABAR Collaboration), 
\Journal{\PRL}{95}{142001}{2005}; K.~Abe \emph{et al} (Belle Collaboration), [hep-ex/0507019];
S.~Uehara \emph{et al} (Belle Collaboration), \Journal{\PRL}{96}{082003}{2006}.
\bibitem{Dsnarrow} B.~Aubert \emph{et al} (BABAR Collaboration), \Journal{\PRL}{90}{242001}{2003}.
\bibitem{DDbarmix} M.~Staric \emph{et al} (Belle Collaboration), \Journal{\PRL}{98}{211803}{2007};
B.~Aubert \emph{et al} (BABAR Collaboration), [hep-ex/0703020].
\bibitem{PDG} W.~M. Yao \emph{et al}, Journal of Physics G {\bf 33}, 1 (2006).
\bibitem{BRSLmc} I.~I. Bigi, M.~Shifman, N.~G.~Uraltsev and A.~Vainshtein, 
\Journal{\PRL}{71}{496}{1993}.
\bibitem{Symanzik} K. Symanzik, \Journal{\NPB}{226}{187}{1983}; \Journal{\NPB}{227}{205}{1983}.
\bibitem{tmQCDsymmetries}
R.~Frezzotti and G.~C.~Rossi,
  JHEP {\bf 0408}, 007 (2004)
  [arXiv:hep-lat/0306014].
\bibitem{AokitmQCD} S. Aoki, \Journal{\PRD}{30}{2653}{1984}.
\bibitem{FGSW} R.~Frezzotti, P.~A.~Grassi, S.~Sint and P.~Weisz  [Alpha Collaboration],
  JHEP {\bf 0108}, 058 (2001)
  [arXiv:hep-lat/0101001].
\bibitem{ETMC} C. Urbach, PoS (LATTICE 2007)022.
\bibitem{ETMCletter} Ph.~Boucaud \emph{et al}, [ETM Collaboration], \Journal{\PLB}{650}{304}{2007} 
[arXiv:hep-lat/0701012].
\bibitem{ZDimopoulos} P. Dimopoulos \emph{et al}, PoS (LATTICE 2007)241.
\bibitem{Dimopscaling} P. Dimopoulos \emph{et al}, PoS (LATTICE 2007)102.
\bibitem{msTarantino} B. Blossier \emph{et al} [ETM Collaboration], [arXiv:0709.4574]; 
V. Lubicz \emph{et al}, PoS (LATTICE 2007)374.
\bibitem{latticemc}  D.~Becirevic, V.~Lubicz and G.~Martinelli,
  Phys.\ Lett.\  B {\bf 524}, 115 (2002)
  [arXiv:hep-ph/0107124];
J.~Rolf and S.~Sint  [ALPHA Collaboration],
  JHEP {\bf 0212}, 007 (2002)
  [arXiv:hep-ph/0209255];
 G.~M.~de Divitiis, M.~Guagnelli, R.~Petronzio, N.~Tantalo and F.~Palombi,
  Nucl.\ Phys.\  B {\bf 675}, 309 (2003)
  [arXiv:hep-lat/0305018];
A.~Dougall, C.~M.~Maynard and C.~McNeile,
  JHEP {\bf 0601}, 171 (2006)
  [arXiv:hep-lat/0508033].
\bibitem{latticefd} A.~Ali Khan {\it et al.}, \Journal{\PLB}{652}{150}{2007} [hep-lat/0701015]; 
G.~M.~de Divitiis, M.~Guagnelli, F.~Palombi, R.~Petronzio and N.~Tantalo,
  Nucl.\ Phys.\  B {\bf 672}, 372 (2003)
  [arXiv:hep-lat/0307005];
 A.~Juttner and J.~Rolf  [ALPHA Collaboration],
  Phys.\ Lett.\  B {\bf 560}, 59 (2003)
  [arXiv:hep-lat/0302016];
A.~Ali Khan {\it et al.}  [CP-PACS Collaboration],
  Phys.\ Rev.\  D {\bf 64}, 034505 (2001)
  [arXiv:hep-lat/0010009];
  C.~Bernard {\it et al.}  [MILC Collaboration],
  Phys.\ Rev.\  D {\bf 66}, 094501 (2002)
  [arXiv:hep-lat/0206016];
C.~Aubin {\it et al.},
  Phys.\ Rev.\ Lett.\  {\bf 95}, 122002 (2005)
  [arXiv:hep-lat/0506030];
E.~Follana {\it et al.}, [arXiv:0706.1726].
\bibitem{CLEOfd} 
M.~Artuso {\it et al.}  [CLEO Collaboration],
  Phys.\ Rev.\ Lett.\  {\bf 95}, 251801 (2005)
  [arXiv:hep-ex/0508057].
\bibitem{CLEOfds} T.K.~Pedlar \emph{et al.}, [arXiv:0704.0439]; 
M.~Artuso \emph{et al.}, [arXiv:0704.0629].
\bibitem{BABARfds} B.~Aubert {\it et al.}  [BABAR Collaboration],
  Phys.\ Rev.\ Lett.\  {\bf 98}, 141801 (2007)
  [arXiv:hep-ex/0607094].
\end{thebibliography}
\end{document}